\documentclass[10pt]{iopart}
\begin{document}
%\preprint{AEI-2000-061,gr-qc/0010007}
\title{Towards the solution of the relativistic gravitational
radiation reaction problem for binary black holes}

\author{Carlos O. Lousto}

\address{Albert-Einstein-Institut, Max-Planck-Institut f{\"u}r
Gravitationsphysik, Am M\"uhlenberg 1, D-14476 Golm, Germany\\
and\\
Instituto de Astronom\'{\i}a y F\'{\i}sica del Espacio--CONICET,
Buenos Aires, Argentina}

\begin{abstract}
Here we present the results of applying the generalized Riemann
$\zeta$-function regularization method to the gravitational radiation reaction
problem. We analyze in detail the headon collision of two nonspinning
black holes with extreme mass ratio. The resulting reaction force
on the smaller hole is repulsive. We discuss the possible extensions
of these method to generic orbits and spinning black holes. The
determination of corrected trajectories allows to add second
perturbative corrections with the consequent increase in the
accuracy of computed waveforms.
\end{abstract}

\pacs{04.30.-w, 04.25.Nx, 04.25.-g, 04.70.Bw}

\vskip 2pc

There has been increasing astronomical evidence
during the last thirty years in favor of the hypothesis that
supermassive black holes, lying in the center of galaxies are 
a common phenomenon in our Universe. They are thought to be
the main engines of quasars, X-ray sources in the core
of galaxy super-clusters, and gamma
ray bursters. There is even compelling evidence that our own
galaxy shelter a black hole of three million solar masses
in its center\cite{bh}.
Although compelling, this is still indirect evidence of
the existence of black holes. It appears that their
characteristic gravitational wave signature provides
the best way to determine its definitive existence.
The LISA project \cite{lisa} offers an excellent opportunity to
match theory and observation in order to reach such goal. From the
characteristics of this space detector, the main astrophysical
sources expected to be "seen" are precisely supermassive black holes
accreting stars close to the galactic center. This scenario offers the
possibility to use perturbation theory around a single
black hole as the appropriate theoretical model to compute
gravitational radiation from such extreme mass ratio binary systems.

Perturbation theory reached maturity
already in the early 1970's\cite{RW57,Z70}. Linearized Einstein
equations have been successfully combined into wave equations for the
two degrees of freedom of the gravitational field propagating
in the curved background of a massive black hole. The
source terms depend on the trajectory of a test
particle moving along a geodesic of the background metric.
There remained one important problem to be solved: The
changes induced in the trajectory of the particle due to
the radiation generated by the orbital motion of the same
particle around the massive hole. Progress along this line
always stumbled with the infinities appearing in the
equation of motion of the particle when perturbations are
evaluated, precisely at the location of the particle.
A consistent regularization method was lacking until recently,
when Mino, Sasaki and Tanaka\cite{MST97} provided a renormalized
equation of motion using two different methods: i)
The Hadamard regularization and ii) An asymptotic matching
of the near and far metric expansions. 
Quinn and Wald \cite{QW97} assume an axiomatic approach to
obtain independently the same equation of motion.
While the above work provide well founded formal expressions,
they appear yet impractical when dealing with concrete
computations since they rely on the use of the harmonic gauge.
Here we present an alternative, practical regularization
method based on the analytically continued Riemann $\zeta-$function.

A test particle with rest mass $m_0$ is here
represented by the energy-momentum tensor
\begin{equation}
T^{\mu \nu }=m_0\frac{U^\mu U^\nu }{U^0r^2}
\delta[r-r_p(t)]\delta^{(2)}[\Omega_p]\ .  \label{tmunu}
\end{equation}

Since the only "forces" acting on the particle are gravitational,
the equation of motion of a test particle is given by the geodesic
equation
\begin{equation}
\frac{d^2x_p^\mu}{d^2\lambda}+\Gamma^\mu_{\alpha\beta}(x_p)\frac{dx_p^\alpha}
{d\lambda}\frac{dx_p^\beta}{d\lambda}=0. \label{geo}
\end{equation}
It is precisely when one wants to evaluate the Christoffel symbols
at the location of the particle that divergences appear. They originate
in the Dirac's delta representation we give to the source term (\ref{tmunu})
of Einstein equations.

In Refs.\ \cite{LP97a,LP97b,LP98} we studied the headon collision of
two Schwarzschild (nonspining) black holes with masses $M$ and $m$
respectively. We treat this problem as perturbations of the Schwarzschild
background. The only expansion parameter being $m/M$, assumed to
be much less than one. The symmetry of the background allows us to
decompose metric perturbations into spherical harmonics
$Y_{\ell m}(\theta,\phi)$ and the gravitational degrees of freedom
satisfy the wave equation
\begin{equation}
-\frac{\partial ^2\psi_\ell }{\partial t^2}+
\frac{\partial ^2\psi_\ell }{\partial r^{*2}}%
-V_\ell (r)\psi_\ell ={\cal S}_\ell (r,t)\ ,  \label{rtzerilli}
\end{equation}
where $r^{*}\equiv r+2M\ln (r/2M-1)$, $V_\ell $ is the Zerilli potential,
and ${\cal S}_\ell (r,t)$, is the source term generated by the small hole
\cite{LP97a,LP97b}.

From the waveform $\psi_\ell(r,t)$ one can reconstruct all perturbed
metric coefficients in the Regge - Wheeler gauge. One can then prove
that all metric coefficients are $C^0$ at $r_p$ for every $\ell$. 
The divergencies reappear when trying to add up {\it all}
multipole contributions \cite{L00}. To see this more explicitly,
let us consider
the $r$ and $t$ components (the only nontrivial for the headon case)
of the geodesic equation (\ref{geo}), and combine them into a single
equation for $r(t)$
\begin{equation}
\frac{d^2r}{dt^2}=\Gamma _{rr}^t\left( \frac{dr}{dt}\right) ^3+\left(
2\Gamma _{tr}^t-\Gamma _{rr}^r\right) \left( \frac{dr}{dt}\right) ^2+\left(
\Gamma _{tt}^t-2\Gamma _{tr}^r\right) \left( \frac{dr}{dt}\right) -\Gamma
_{tt}^r.  \label{geo1}
\end{equation}
Linearization of this equation and subtraction of the zeroth order
geodesic gives the deviation of the trajectory $\Delta r_p(t)$ from the
zeroth order one
\begin{eqnarray}
\dot r_p(t)=\partial_t z_p=-(1-2M/z_p)\sqrt{\frac{2M/z_p-2M/z_0}{1-2M/z_0}},
\end{eqnarray}
directly in terms of Schwarzschild coordinates, ready for further applications
\begin{eqnarray}
\Delta\ddot r_p=A\ \Delta r_p+B\ \Delta\dot r_p+C
\end{eqnarray}
where
\begin{eqnarray}\label{C}
A&=&\frac{2M}{r_p^3}\left[3-\frac{3M}{r_p}
-\frac{(1-3M/r_p)\dot r_p^2}{(1-2M/r_p)^{2}}\right],\nonumber\\ 
B&=&\frac{6M\dot r_p}{r_p^2(1-2M/r_p)},\nonumber\\
C&=&\left[
\stackrel{(1)}{\Gamma^t}_{rr}\dot r_p ^3+
\big(2\stackrel{(1)}{\Gamma^t} _{tr}-\stackrel{(1)}{\Gamma^r} _{rr}\big)
 \dot r_p ^2+\big(\stackrel{(1)}{\Gamma^t} _{tt}-
2\stackrel{(1)}{\Gamma^r} _{tr}\big) \dot r_p
-\stackrel{(1)}{\Gamma^r}_{tt}\right]\nonumber\\
&=&\sum_{l=0}^\infty C_\ell .
\end{eqnarray}
It is precisely $C_\ell$ the piece that being finite for each multipole
diverges when summed over $\ell$ (here we consider the averaged value
of $C_\ell$ over both radial sides of $r_p$). The direct
numerical integration of the wave equation (\ref{rtzerilli})
shows that we can split $C_\ell$ as
\begin{eqnarray}
C&=&\sum_{\ell=0}^\infty\left\{(2\ell+3/2-\beta)^{-\beta+1/2} C_\infty +
C_{\ell}^{ren}\right\},
\end{eqnarray}
where the introduction of $\beta$ is motivated by the $D$-dimensional
extension of the initial value problem for time symmetric conformally
flat data. In this case $-\beta=D-9/2$. 
We neatly determined numerically that $\beta=1/2$ along the
trajectory of the particle.

This form of the coefficient $C$, determining the corrected trajectory
of the particle, shows clearly that the $\ell-$independent term, summed
over $\ell$, diverges. We also determined numerically that $C_{\ell}^{ren}$
behaves like $\ell^{-2}$ for large $\ell$, thus giving raise to
a finite contribution to $C$. The key observation here is that we
can bring the divergent sum to the form of the Riemann $\zeta-$ function
\cite{BD82}, 
$\zeta(a,b)=\sum_{\ell=0}^\infty(\ell+b)^{-a}$,
\begin{eqnarray}
C=2^{-\beta+1/2}C_\infty\zeta(\beta-1/2,1/2)
+\sum_{\ell=0}^\infty C_\ell^{ren}.
\end{eqnarray}
Since the analytically continued
$\zeta$-function gives $\zeta(0,1/2)=0$, in order to regularize $C$,
we must just subtract to each multipole their $\ell\to\infty$ piece.
When we do that \cite{L00} we have found that the effect of $C^{ren}$ on the
trajectory is to generate a repulsive contribution that has a maximum
near the pick of the Zerilli potential at $r=3.1M$.

Let us label functions with the $\pm$ superindex to refer to their
values in the region $r>r_p$ and $r<r_p$ near the location of the
particle $r_p$ respectively. One can see that both, initial data and
Zerilli equation imply the following formal
symmetry% in the waveform $\psi_\ell$
\begin{eqnarray}
\sqrt{L}\psi^{\pm}_\ell(-L)=\sqrt{L}\psi^{\mp}_\ell(L)
\end{eqnarray}
where $L=\ell+1/2$. The same symmetry holds for the $t$ and $r$ derivatives,
hence the reaction `force' will have the form
${\cal F}^{\pm}({L}) = p({L}^2)\pm {L}\ q({L}^2)$, and
the large ${L}$ expansion of its average
\begin{eqnarray}
<{\cal F}_p>&=&{\sum_{\ell=0}^{\infty}}<{\cal F}_\ell>={\sum_{\ell=0}^{\infty}}p({L}^2)={\sum_{\ell=0}^{\infty}}\sum_{n=0}^{\infty}\frac{A_n}{{L}^{2n}}
<{\cal F}_p>\nonumber\\
&=&\sum_{n=0}^{\infty}A_n{\sum_{\ell=0}^{\infty}}{{L}^{-2n}}
=\sum_{n=0}^{\infty}A_n\ {\zeta}(2n,1/2)\nonumber\\
&=&A_0\ {\zeta}(0,1/2)+A_2\ {\zeta}(2,1/2)+A_4\ {\zeta}(4,1/2)+\ldots
\end{eqnarray}
where we observe that ${\zeta}(0,1/2)=0$ gives the regularization,
${\zeta}(2,1/2)=\frac12\,{\pi }^{2}$ gives the leading reaction `force'
and ${\zeta}(4,1/2)=\frac16\,{\pi }^{4}$ gives an estimate of the
`{\it error}' of the leading term of the order of $7-8\%$.

Barack, Ori and Burko \cite{BO00,BB00} have developed independently
a similar approach (in the sense that they also decompose into
multipoles), but that uses the regularization method of Ref.\ \cite{MST97}
and implements it to scalar radiation in the Schwarzschild background.
Their results are completely compatible with those of the
$\zeta-$function regularization in the sense that in the end it is
exactly only the $\ell\to\infty$ that has to be subtracted to the
originally divergent force.

With Barack \cite{BL01} we have been able to confirm analytically
the numerical behavior discussed above. Based on a local analysis
of the two-point Green's function we make a $1/L$ (with $L=\ell+1/2$)
expansion to determine the behavior of the waveform $\psi_\ell$ and its
derivatives for large $\ell$, for instance
\begin{eqnarray}
\psi_\ell^\pm(r_p)&=&4m_0\sqrt{2\pi}\sqrt{L}
\left\{L^{-3}\pm2EL^{-4}+{\cal O}(L^{-5})\right\}\nonumber
\\
\partial_r\psi_\ell^\pm(r_p)&=&\frac{4m_0\sqrt{2\pi}}{r_p-2M}\sqrt{L}
\left\{\mp EL^{-2}-\frac32E^2L^{-3}\right.\nonumber
\\
&&\pm\left.\left(\frac{6M}{r_p}-\frac94\right)
EL^{-4}+{\cal O}(L^{-5})\right\}\nonumber\\
\partial_t\psi_\ell^\pm(r_p)&=&-(1-2M/r_p)\left(\frac{\dot{r}_p}{E}\right)
\partial_r\psi_\ell^\pm(r_p)
\end{eqnarray}
for a particle released from rest at $r_0$, $E=\sqrt{1-2M/r_0}$.
$\pm$ stands for the
side derivatives about $r_p$ (waveforms are discontinuous).
Hence we can compute
\begin{equation}
C_\ell^{\pm}=a^{\pm}L+b+c^{\pm}L^{-1}+O(L^{-2}),
\end{equation}
where $a^{\pm}$, $b$, and $c^{\pm}$ are $l$-independent coefficients.
We find that the average value of $a$ and $c$ vanish at the location
of the particle, while
%the asymptotic behaviour of
the {\em average}
%$\tilde C\equiv \sqrt{2l+1}\,C$ is
\begin{equation}
<C(l\to\infty)>=
-\frac{\sqrt{\pi}\,m_0}{r^2}(1-2M/r_p)^{3/2}E^{-2}
\end{equation}

This method can be used to assist the numerical computation for large $\ell$
carrying out the expansion to order $1/L^2$ in $C$.
Thus, making necessary the numerical integration for only the first few
lower multipoles. This might be crucial when dealing with
generic orbits around Schwarzschild black holes. The analytic expansions
would allow a clean application of the $\zeta-$regularization scheme and
sum over large $\ell$ of the regularized reaction "force".

In the orbital case it will also be important to test our final
results with those of the energy-balance approximation
\cite{CKP94,H00,FK00}.
This method provides accurate results for circular orbits on
the Kerr background \cite{H00}. Our method, in principle, could also
be implemented to a particle orbiting a Kerr hole since the Teukolsky
equation, that describes the perturbations around rotating holes, can
be decomposed into multipoles in the frequency domain (Laplace decomposition
of the time-dependence \cite{CL98}).
The details of this implementation remain to be explicitly worked out.

A further application of the above results one can foresee is the
extension of the analysis to second perturbative order \cite{CL99}.
In this way
one can increase the accuracy with which waveforms are computed
as well as reaching not so small mass ratios $m/M$. At this stage,
the resulting waveforms will not only be relevant for LISA,
but also for ground based gravitational wave detectors sensible to
galactic binary black holes of comparable masses and black hole -
neutron star systems.

I would like to end this contribution with an optimistic note
from the theoretical side and predict that the subsequent
progress in our understanding of radiation reaction
will soon bring good news for modeling gravitational emission
from astrophysical sources.

%\vskip 0.3cm
\medskip
%\ackn
{\it Acknowledgments}: Thanks to L. Barack for many useful remarks.

%\newpage
\appendix

\section{Connection coefficients}

Here we make explicit the details that enter into the computation
of $C_\ell$ given in Eq.\ (\ref{C}). The connection coefficients
involved in the first order geodesic, Eq. (\ref{geo1}) in terms
of the first order perturbations as defined in the Regge - Wheeler
(RW) gauge \cite{RW57} are
\begin{eqnarray}
\Gamma _{tt}^t &=&\frac M{r^2}H_1-\frac 12\stackrel{.}{H}_0,\qquad \Gamma
_{tr}^t=\frac M{r^2}\left( 1-\frac{2M}r\right) ^{-1}-\frac 12H_0^{\prime }, 
\nonumber \\
\Gamma _{rr}^t &=&\left( 1-\frac{2M}r\right) ^{-2}\left[ \frac 12\stackrel{.%
}{H}_2-\left( 1-\frac{2M}r\right) H_1^{\prime }-\frac M{r^2}H_1\right] ,
\label{conecciones} \\
\Gamma _{tt}^r &=&\left( 1-\frac{2M}r\right) \left[ \stackrel{.}{H}_1+\frac M%
{r^2}\left( 1-H_2-H_0\right) -\frac 12\left( 1-\frac{2M}r\right) H_0^{\prime
}\right] ,  \nonumber \\
\Gamma _{tr}^r &=&-\frac M{r^2}H_1+\frac 12\stackrel{.}{H}_2,\qquad \Gamma
_{rr}^r=-\frac M{r^2}\left( 1-\frac{2M}r\right) ^{-1}+\frac 12H_2^{\prime },
\nonumber
\end{eqnarray}
where the overdot and prime mean derivative with respect to the
Schwarzschild time $t$ and radial coordinate $r$ respectively.
In addition, we have from Einstein equations in the RW gauge
(see Ref. \cite{Z70}, Eq. (C7g)), that for a radial infall 
\begin{equation}
H_0=H_2,\quad  \ell\geq 2.\label{lapso}
\end{equation}
This allows us to write everything in terms of $H_1^\ell$ and $H_2^\ell$
only (and its derivatives). $K^\ell$ does not appears in the headon 
geodesics.

%\section{Metric Derivatives}

Proven the continuity of the metic coefficients at $r=r_p$ our way is clear
to compute the trajectory of the particle to first perturbative order.
Our last step is to compute these metric derivatives in terms of $%
\psi $ and $\stackrel{.}{\psi }$.
Since we know that the metric coefficients are continuous, their derivatives
will have a jump, but not delta (or derivatives of it) terms. Then, in order
to simplify expressions we will not write the source terms and use an
overbar $(\stackrel{\_\_}{\psi })$
on radial derivatives of $\psi $ to indicate that we have already
subtracted its singular behavior at $r=r_p$ and taken the average value.

The $r$ - derivative of $H_2$ from Ref.\ \cite{L00} is
\begin{eqnarray}
&&\partial _rH_2 =\left( r-2M\right) \partial _r^3\stackrel{\_\_}{\psi }%
+\left[ 1+\frac{\lambda r^2-\lambda Mr+3M^2}{(\lambda r+3M)r}\right]
\partial _r^2\stackrel{\_\_}{\psi }  \nonumber \\
&+&\ \frac{-\lambda ^2\left( \lambda +1\right) r^3-\lambda \left( 2\lambda
-3\right) Mr^2-15\lambda M^2r-18M^3}{r^2(\lambda r+3M)^2}\partial _r%
\stackrel{\_\_}{\psi }  \label{primaH2} \\
&+&\frac{\lambda ^3\left( \lambda +1\right) r^4+3\lambda ^2\left( \lambda
-1\right) Mr^3+27\lambda ^2M^2r^2+63\lambda M^3r+54M^4}{r^3\left( \lambda
r+3M\right) ^3}\stackrel{\_\_}{\psi }  \nonumber
\end{eqnarray}

Likewise, upon $r$ - derivation of expression the expression for
$H_1$, given in Ref.\ \cite{L00} we obtain 
\begin{eqnarray}
\partial _rH_1 &=&r\partial _r^2\stackrel{.}{\stackrel{\_\_}{\psi }}+\frac{%
2\lambda r^2+\left( 3-5\lambda \right) Mr-9M^2}{\left( r-2M\right) (\lambda
r+3M)}\partial _r\stackrel{.}{\stackrel{\_\_}{\psi }}  \nonumber \\
&&\ \ +\frac{\left[ \lambda \left( \lambda +3\right) r^2-6\lambda Mr+3\left(
4\lambda +3\right) M^2\right] M}{\left( r-2M\right) ^2(\lambda r+3M)^2}%
\stackrel{.}{\stackrel{\_\_}{\psi }}.
\end{eqnarray}

The time derivative of $H_2$ can be also obtained% from Eq. (\ref{H2extract})

\begin{eqnarray}
\partial _tH_2&=&(r-2M)\partial _r^2\stackrel{.}{\stackrel{\_\_}{\psi }}+%
\frac{3M^2-\lambda Mr+\lambda r^2}{r(\lambda r+3M)}\partial _r\stackrel{.}{%
\stackrel{\_\_}{\psi }}  \nonumber \\
&&\ -\frac{9M^3+9\lambda M^2r+3\lambda ^2Mr^2+\lambda ^2(\lambda +1)r^3}{%
r^2(\lambda r+3M)^2}\,\stackrel{.}{\stackrel{\_\_}{\psi }}
\end{eqnarray}

and finally we find $\partial _tH_1$ from Zerilli's\cite{Z70} Eq.\ (C7e),
where $K^{\prime }$ can be found from $H_2$, given in Ref.\ \cite{L00},
%Eq. (\ref{H2Kextract}),
and $H_2^{\prime }$ is given by Eq. (\ref{primaH2}), 
%and $H_2$ by Eq. (\ref{H2extract}) 
\begin{eqnarray}
&&\partial _tH_1=\frac{\left(r-2M\right) ^2}r\partial_r^3
\stackrel{\_\_}{\psi }+\left(r-2M\right)
\,\frac{\left( 15\,M^2+3\,\lambda \,M\,r+\lambda \,r^2\right) }{r^2\,\left(
3\,M+\lambda \,r\right) }\partial _r^2\stackrel{\_\_}{\psi }\nonumber \\
&-&2\,\left(2\,{\lambda}^{2}M{r}^{3}+18\,\lambda\,{M}^{2}{r}^{2
}+27\,r{M}^{3}-9\,{\lambda}^{2}{M}^{2}{r}^{2}-42\,\lambda\,{M}^{3}r
\right.\nonumber\\
&-&\left.63
\,{M}^{4}+{\lambda}^{3}{r}^{4}+{\lambda}^{2}{r}^{4}-2\,M{\lambda}^{3}{
r}^{3}\right)/\left ({r}^{3}\left (\lambda\,r+3\,M\right )^{2}\right )
{\partial_r }\stackrel{\_\_}{\psi}\nonumber\\
&+&2\left( -189\,M^5+81\,M^4\,r-216\,\lambda
\,M^4\,r+90\,\lambda \,M^3\,r^2-90\,\lambda ^2\,M^3\,r^2\right. \nonumber\\
&+&\left.33\lambda^2M^2r^3
-18\lambda ^3M^2r^3-3\lambda^3(\lambda-1)Mr^4+\lambda
^3r^5+\lambda ^4r^5\right)\stackrel{\_\_}{\psi }\nonumber\\
&& /\left( r^4\left( 3M+\lambda r\right)
^3\right)
\end{eqnarray}

These expression complete the equations needed to (numerically) integrate
the geodesic equation (\ref{geo1}) that gives us the particle's trajectory
to first perturbative order. Note that the numerical implementation should
be able to handle third order derivatives of $\psi_\ell$.

\end{document}